\pdfoutput=1

\documentclass[11pt]{article}

\usepackage[preprint]{acl}

\usepackage{times}
\usepackage{latexsym}

\usepackage[T1]{fontenc}

\usepackage[utf8]{inputenc}

\usepackage{microtype}

\usepackage{inconsolata}

\usepackage{graphicx}
\usepackage{multirow}

\title{GR-LLMs: Recent Advances in Generative Recommendation Based on Large Language Models}

\author{Zhen Yang \quad \textbf{Haitao Lin}  \quad Jiawei Xue \quad Ziji Zhang \\ 
    AMAP, Alibaba Group\\
    \texttt{\{zhongming.yz,lht416932,xuejiawei.xjw,zhangziji.zzj\}@alibaba-inc.com}}
    
\begin{document}
\maketitle
\begin{abstract}
In the past year, Generative Recommendations (GRs) have undergone substantial advancements, especially in leveraging the powerful sequence modeling and reasoning capabilities of Large Language Models (LLMs) to enhance overall recommendation performance.  LLM-based GRs are forming a new paradigm that is distinctly different from discriminative recommendations, showing strong potential to replace traditional recommendation systems that are heavily dependent on complex, hand-crafted features. In this paper, we provide a comprehensive survey designed to facilitate further research on LLM-based GRs. Initially, we outline the general preliminaries and application cases of LLM-based GRs. Subsequently, we introduce the main considerations during the industrial applications of GRs. Finally, we explore promising directions for LLM-based GRs.  We hope that this survey contributes to the ongoing advancement of the GR domain.

\end{abstract}

\section{Introduction}
Recommendation systems \cite{adomavicius2005toward,li2024survey}, which aim to recommend the items (e.g., e-commerce products, micro-videos, news, and point-of-interests) by implicitly inferring user interest from the user's profile and historical interactions, are ubiquitous in the modern digital landscape, serving as critical interfaces for navigating the vast sea of information and choices available online. The effectiveness of recommendation systems has been a driving force behind the success of numerous online platforms, from e-commerce giants and social networks to content streaming services and news aggregators. 

With the advancement of recommendation systems, modeling algorithms have roughly undergone three different technological paradigms, namely \textbf{machine learning-based recommendation (MLR)}, \textbf{deep learning-based recommendation (DLR)}, and 
\textbf{generative recommendation (GR)}.  The \textbf{MLR} primarily relies on traditional machine learning algorithms, often built upon explicit feature engineering. Key techniques include collaborative filtering \cite{breese2013empirical,he2017neural,sarwar2001item,linden2003amazon}, which predicts user preferences based on similarity with other users or items, and content-based filtering, which recommends items similar to those a user has liked based on item attributes. Matrix factorization techniques \cite{koren2009matrix,rendle2012bpr}, such as singular value decomposition or alternating least squares, are also central to this era, aiming to learn latent factors that represent user and item preferences to predict missing entries in a user-item interaction matrix. While \textbf{MLR} focuses on capturing statistical patterns in historical data, it frequently faces challenges in addressing data sparsity and the cold-start problem (where new users or items lack sufficient interaction data). Feature engineering remains essential for providing meaningful input features. The \textbf{DLR} leverages the power of deep neural networks to automatically learn complex, non-linear representations directly from raw or sparse features \cite{tang2018personalized,xue2017deep,zhang2017joint,chen2019dynamic, xue2025hgcl}. In industrial recommendation systems, \textbf{DLR} has been used for nearly a decade, typically with inputs that include many well-designed handcrafted features to improve model performance. The primary challenge with \textbf{DLR} models lies in the trade-off between effectiveness and efficiency. Their relatively small parameter size often makes scaling more difficult, hindering convenient increases in model capacity to enhance recommendation quality. 

Traditional recommendation paradigms, i.e., the \textbf{MLR} and \textbf{DLR}, focus on predicting a similarity or rank score based on hand-crafted feature engineering and intricate cascaded modeling structure,  making them brittle, difficult to interpret, and requiring significant manual effort for maintenance and adaptation to new data or domains. 
In recent years, a paradigm shift has been catalyzed by the emergence and rapid development of large language models (LLMs). These models, exemplified by architectures like GPT, BERT, encoder-decoder Transformer, and others, trained on vast amounts of text data, demonstrate remarkable capabilities in understanding and generating human language \cite{floridi2020gpt,liu2019roberta,guo2025deepseek}. They excel at capturing complex statistical dependencies in sequential data, performing sophisticated reasoning tasks, and exhibiting a deep understanding of context and semantics. This unprecedented power in sequence modeling and general-purpose reasoning has naturally led to the emergence and rapid advancement of \textbf{GRs}. In particular, the \textbf{GR} systems have seen great progress in the past year. SASRec \cite{kang2018self} first proposes to predict the next user-interacted item through an autoregressive approach based on a transformer model. \citet{zhai2024actions} propose HSTU, a new transformer framework for better modeling sequences and inference efficiency, which is followed by \citet{huang2025towards} on ranking tasks. To deal with a large number of items in industrial-scale recommendation, TIGER \cite{rajput2023recommender} incorporates the idea of RQ-VAE \cite{zeghidour2021soundstream} to learn to transform items into multiple semantic IDs, largely reducing the vocabulary size.  Building upon similar semantic ID encoding, OneRec \cite{deng2025onerec} employs a Mixture of Experts (MoE) architecture \cite{dai2024deepseekmoe} and a Direct Preference Optimization (DPO) strategy \cite{rafailov2023direct} to further improve the recommendation ability. For industrial-scale generative recommendation, MTGR \cite{han2025mtgr} proposes incorporating the cross features used in \textbf{DLR} and finds that excluding cross features severely damages the model's performance. \citet{qiu2025one} and \citet{zheng2025beyond} further propose an end-to-end generative architecture that unifies online advertising ranking as one model. Moreover, \citet{jiang2025large} points out that the LLM can be used as a universal recommendation learner, and they propose URM, which can perform well on versatile recommendation tasks. 

Works mentioned above show that LLM-based GRs represent a fundamental departure from traditional discriminative methods. This generative aspect offers several compelling advantages. Firstly, it allows for greater explainability, enabling systems to communicate why a particular item is recommended, thereby building user trust and facilitating feedback loops. Secondly, it inherently supports creativity and novelty, as LLMs can suggest items beyond the most predictable ones based on past behavior, potentially helping users discover new interests. Thirdly, the unified language model approach potentially simplifies system design by reducing the need for complex, hand-crafted feature engineering and separate modules for different tasks. Furthermore, the scaling laws in LLMs have great potential to raise the performance ceiling of generative recommendation systems.

Recognizing the transformative potential and the rapid pace of development in this area, there is a growing need for a comprehensive survey that synthesizes the current knowledge on GRs. While initial explorations into this space have been conducted, the field is evolving rapidly, and a systematic survey is crucial to help researchers and practitioners navigate the landscape, understand the core concepts and techniques, learn from existing applications, and identify promising avenues for future work. In this paper, we provide a comprehensive survey aimed at facilitating further research and development in LLM-based GRs. We structure our survey to first outline the general preliminaries and foundational concepts of LLM-based GRs. Subsequently, we delve into the diverse application cases and real-world deployments of these systems. Finally, we critically analyze the main considerations and challenges encountered when applying LLM-based GRs in demanding industrial scenarios. We conclude by exploring promising future research directions. We hope that this survey contributes significantly to the ongoing advancement and maturation of the GR domain.

\section{Preliminaries}
\subsection{Large language models}
Large language models, which are trained on vast amounts of text data, have demonstrated significant capabilities in natural language processing \cite{bubeck2023sparks,yang2025qwen3,grattafiori2024llama}. Given an input sequence $X=\{x_1, x_2, \ldots, x_n\}$, LLMs are trained to optimize the probability $P(x_t|x_{<t};\theta)$ with the next-token prediction format, where $\theta$ represents the parameters of the model and $x_{<t}$ indicates the tokens before $x_t$.
Initially, LLMs were primarily text-based, but they have evolved to handle multi-modal data, integrating text with images, audio, and video \cite{team2023gemini,liu2023visual,liu2023improved,yang2023teal}. With the ability to support multi-modal inputs and outputs, large models can perform a variety of sequence generation tasks. 

\subsection{Traditional cascaded recommendations}
\begin{figure}[t]
    \centering
    \includegraphics[width=0.7\hsize]{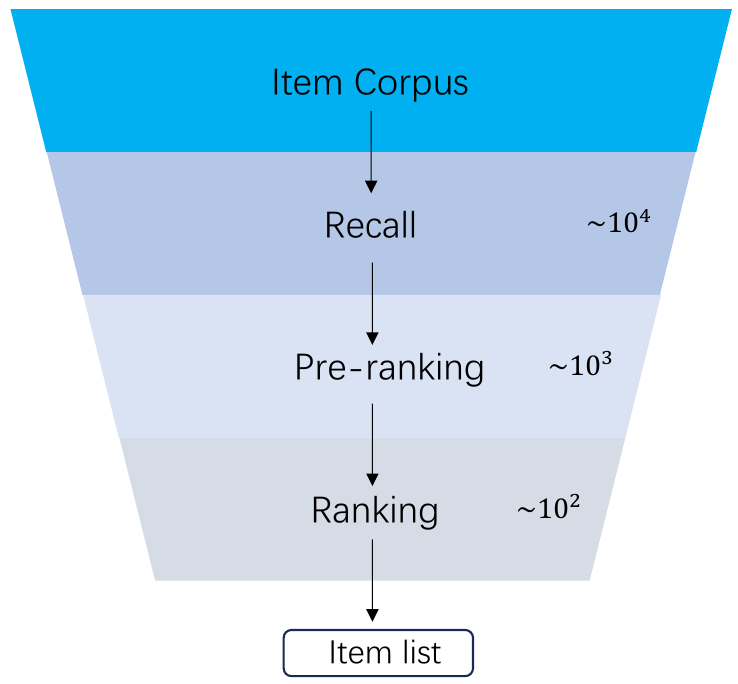}
    \vspace{-1mm}
    \caption{A typical cascade ranking system, which includes three stages from the top to the bottom: Recall, Pre-ranking, and Ranking.}
    \label{cascade}
\end{figure}

Traditional recommendation systems widely adopt multi-stage cascaded architectures to balance computational efficiency and prediction accuracy \cite{burges2010ranknet,chang2023twin,wang2011cascade}. As illustrated in Figure \ref{cascade}, a typical cascaded recommendation system includes three sequential stages: recall, pre-ranking, and ranking. Although efficient in practice, existing methods typically treat each stage independently, where the effectiveness of each isolated stage serves as the upper bound for the subsequent stage, thereby limiting the performance of the overall recommendation system. Many previous works \cite{fei2021gemnn,gallagher2019joint,huang2023cooperative,wang2024adaptive} have been proposed to enhance overall recommendation performance by enabling interaction among different stages, but they still maintain the traditional cascade paradigm. Recently, the GRs have emerged as a promising paradigm to serve as a unified architecture for end-to-end generation \cite{qiu2025one,deng2025onerec}.

\section{{Application Settings of GR}}
In the past year, various GR systems have achieved significant business benefits in practical industrial settings. There are two different branches of approaches applying GRs in online recommendation: the first branch is cooperating with the corresponding modules of traditional cascaded systems; the other is to apply generative models directly for end-to-end recommendations. This section will systematically summarize and analyze recent works based on their specific application settings.

\subsection{Recall}
Recall is a foundational step that narrows the candidate item pool to a subset potentially relevant to a user. This stage is critical for supporting the subsequent ranking stage and prioritizes efficient implementation over massive datasets \cite{julianbook}. 
LLMs can be leveraged in the recall phase in three approaches: prompt-based, token-based, and embedding-based recall methods.

Prompt-based methods generate recall results by providing user information to pre-trained LLMs through customized queries. Pre-trained LLMs, such as Qwen3 \cite{yang2025qwen3}, are equipped with extensive world knowledge, which enables the recall of items even when user information is implicit or sparse. Specifically, LLMTreeRec \cite{zhang2024llmtreerec} constructs prompts to guide LLMs in summarizing user interests, inferring related item categories, and recalling specific items through item trees. Note that the parameters in LLMs remain frozen and are not fine-tuned in this method. Besides, SyNeg~\cite{li2024syneg} utilizes LLMs to synthesize negative samples that are difficult to classify. These generated hard negative samples are then fused with the retrieved negatives for model fine-tuning.

Token-based methods map user behavior sequences to token sequences, formulating the recall problem as the next token prediction task. These methods offer strong flexibility in token sequence construction, model architecture design, and information encoding and decoding, supporting various industrial deployment scenarios. Notably, HSTU \cite{zhai2024actions} reformulates the recall task within a generative framework. 
Similarly, KuaiFormer \cite{liu2024kuaiformer} defines the recall process as next token prediction, incorporating hierarchical user behavior summarization across early, middle, and recent action sequences. The model has been deployed to industrial-scale platforms serving 400 million daily active users with a significant lift in both offline and online evaluation metrics. To unify diverse recall objectives (e.g., items users will click on), Universal Retrieval Model (URM) \cite{jiang2025large} encodes these objectives as components within token sequences to provide objective-aware recall outcomes. 

Embedding-based methods leverage LLMs as encoders to generate item embeddings, which are subsequently integrated into classical DLR methods. For example, MoRec employs pre-trained vision and text encoders to obtain item embeddings~\cite{yuan2023go} for downstream prediction modules such as DSSM~\cite{huang2013learning} and SASRec~\cite{kang2018self}.

\subsection{Rank}
The ranking stage plays a pivotal role in determining the final relevance and diversity of candidate items presented to users.
Compared to DLR, LLMs can accurately model user preferences mainly through chronological user behavior sequences, which eliminates the need for extensive feature engineering. Furthermore, LLMs possess a theoretical foundation in scaling laws, which allows LLM-based ranking systems to overcome the performance bottlenecks inherent in DLR approaches by leveraging increased model scale \cite{zhai2024actions}.

Current approaches to integrating LLMs into the ranking stage can be categorized into two paradigms: generative architectures and hybrid integration architectures, depending on whether they retain the traditional DLR framework. Generative architectures abandon DLR frameworks, employing LLMs to directly process user behavior sequences and generate candidate scores through explicitly supervised tasks such as CTR prediction. As mentioned above, GR \cite{zhai2024actions} constructs chronological sequences by integrating user-related behaviors and features, reformulating the ranking task as a sequential transduction task. Notably, GR marks the first observation of the scaling laws inherent in LLMs within large-scale recommendation systems. GenRank \cite{huang2025towards} proposes an action-oriented sequence organization that treats items as positional context and focuses on predicting user actions associated with each candidate item. DFGR \cite{guo2025action} introduces a dual-flow generative architecture that decouples user behavior sequences into parallel real and fake flows to address computational inefficiencies in GR. It merges real and fake action-type tokens to model heterogeneous user behaviors while maintaining end-to-end efficiency. 

Hybrid integration architectures, in contrast, leverage LLMs to generate highly informative representations that are integrated as supplementary features to DLR, thereby enhancing the performance of existing systems. As the representative approach, LEARN \cite{jia2025learn} adapts a fixed LLM as the Content-Embedding Generation (CEG) module to preserve open-world knowledge while bridging the gap between general and collaborative domains via a twin-tower structure comprising a user tower and an item tower. LEARN deploys its LLM-derived representations as complementary features within existing ranking models in industrial scenarios, achieving substantial improvements in practical applications. HLLM \cite{chen2024hllm} employs a hierarchical LLM architecture that sequentially employs a trainable ITEM LLM and USER LLM to extract item-specific and user-specific representations, respectively. In online scenarios, HLLM integrates the high-level item and user representations through a late fusion approach. SRP4CTR \cite{han2024enhancing} enhances CTR prediction by integrating self-supervised sequential recommendation pre-training with CTR models through a fine-tuned architecture that employs FG-BERT for multi-attribute side information encoding and a uni cross-attention block to transfer knowledge from pre-trained sequences to item-specific predictions efficiently. To integrate the strengths of both paradigms, MTGR \cite{han2025mtgr} employs a generative architecture based on HSTU to model user-level data while retaining raw features, including cross features designed for DLR.

\subsection{End-to-end recommendation}
The goal of end-to-end recommendation is to directly output the recommendation results based on users' historical behaviors. A key distinction from the recall task lies in whether the output inherently possesses ranking capability and whether it can fully replace the conventional cascaded recommendation pipeline. The main advantage of end-to-end recommendation is that it could avoid error propagation and objective misalignment caused by pipeline-based methods.

Most recall-focused methods (e.g., TIGER \cite{rajput2023recommender}, COBRA\cite{yang2025sparsemeetsdenseunified} and URM \cite{jiang2025large}) are trained with the objective of next item prediction, emphasizing top-k hit rate during evaluation. In contrast, end-to-end approaches often enhance the model’s ranking ability through additional post-training stages built upon generative models. OneRec \cite{deng2025onerec} first deploys the end-to-end recommendation model to the industrial scenario, targeting the generation of session-wise recommendation lists as its core task. It employs an iterative reward-model-based DPO alignment strategy to progressively refine the model. It completely replaces the cascaded online recommendation pipeline with a single model, achieving significant improvements on online feedback metrics. OneSug \cite{guo2025onesug} employs list-wise preference alignment, leveraging online feedback signals for reinforcement training to generate more preferable query suggestions. EGA-V2 \cite{zheng2025beyond} is specifically optimized for advertising recommendation scenarios, where it trains a reward model on top of next item prediction to evaluate the expected cumulative reward of an ad sequence, thereby improving the overall expected return of recommended ads. Although EGA-V1 \cite{qiu2025one} is primarily designed as a ranking model, it considers the full set of candidate advertisements as input, which places it within the scope of end-to-end recommendation.

Beyond industrial applications, several academic studies have focused on how to learn better ranking capabilities within the end-to-end framework. S-DPO \cite{chen2024softmax} is the first work to apply the DPO algorithm to recommendation models. By considering multiple negative samples exposed at the same time, S-DPO extends the binary preference optimization of DPO into a multi-preference setting, thereby improving the ranking quality of recommendations. RosePO \cite{liao2024rosepo} further enhances the construction of negative samples by incorporating factors such as popularity and semantic similarity, generating harder negatives for more effective reinforcement learning. SPRec \cite{gao2025sprec} introduces a self-replay mechanism, using the model's predicted results as negative samples, which increases the difficulty of distinguishing between positive and negative examples and leads to stronger generalization. Combining these reinforcement learning strategies with industrial applications represents a promising direction for the development of end-to-end recommendations.

\section{Main Considerations and Challenges}
As a new paradigm in recommendations, GR faces many challenges in industrial applications; this section will detail the key issues that need to be considered during the application process.
\subsection{Training pipelines}
%
\begin{table*}[htbp]
\centering
\caption{A summary of learning targets and post-training strategies of recent GR works.}
\label{tab:training pipelines}
\resizebox{0.9\textwidth}{!}{
\begin{tabular}{l|ccc}
\hline
                           & Methods    & Training Objective                & Training Strategy \& Loss         \\ \hline
\multirow{4}{*}{Single Stage} & HSTU \cite{zhai2024actions}       & next item / action prediction    &    cross entropy                                        \\
                           & URM \cite{jiang2025large}       & next item prediction           &     cross entropy                                  \\
                           & KuaiFormer \cite{liu2024kuaiformer} & user-item embedding similarity &      in-batch contrastive learning                                         \\
                           & MTGR \cite{han2025mtgr}       & ctr prediction                 &                   binary classification                                  \\ \hline
\multirow{8}{*}{Multi Stage}               & OneRec \cite{deng2025onerec}     & next item prediction           & DPO loss, iterative alignment                      \\
                           & OneSug \cite{guo2025onesug}    & next item prediction           & DPO loss, list-wise preference alignment           \\
                           & EGA-V2 \cite{zheng2025beyond}    & next item prediction           & auction-based preference alignment                 \\
                           & EGA-V1 \cite{qiu2025one}    & ctr prediction                 & auction-based preference alignment       \\
                           & QARM \cite{luo2024qarm}  &  item alignment  & contrastive learning, id as ranking feature \\
                           & HLLM  \cite{chen2024hllm}     & ctr prediction                 & contrastive learning, embedding as ranking feature \\
                           & LEARN \cite{jia2025learn}     & ctr prediction                 & contrastive learning, embedding as ranking feature \\
                           & LUM  \cite{yan2025unlocking}      & next item prediction           & contrastive learning, embedding as ranking feature \\ \hline
\end{tabular}
}
\end{table*}

A core issue in GR is how to design the training methodology and objectives to align with the recommendation task. Based on the number of stages involved in the training process, we categorize existing approaches into two main types: single-stage training and multi-stage training. Table~\ref{tab:training pipelines} summarizes these methods by category, training objective, and distinctive strategies.

\subsubsection{Single-stage training}
In single-stage training, the model produces the final recommendation output through a single training phase, typically focusing on only one specific task—either recall or ranking. For recall tasks, a common training objective is next item prediction. Methods such as HSTU \cite{zhai2024actions} and URM \cite{jiang2025large} train models based on this objective, aiming to predict the top-k items as the recall results. Given the large scale of item spaces, these approaches often employ negative sampling during training to reduce the computational cost of the softmax operation. In contrast, KuaiFormer \cite{liu2024kuaiformer} adopts in-batch softmax, generating embeddings directly for recall purposes. For ranking tasks, a typical training objective is CTR prediction. MTGR \cite{han2025mtgr} explicitly models the relationship between user sequences and candidate items, predicting CTR values to replace traditional ranking models, thereby improving the recommendation performance.

\subsubsection{Multi-stage training}
Multi-stage training generally involves two phases: pre-training and fine-tuning. By defining distinct training objectives at each stage, the model can learn different capabilities. This category can be further divided into two subcategories, depending on how the pretrained models are utilized during fine-tuning: representation-based fine-tuning and model-based fine-tuning.

\paragraph{Representation-based finetuning}
Methods in this category (e.g., HLLM \cite{chen2024hllm}, LEARN \cite{jia2025learn}, and LUM \cite{yan2025unlocking}) primarily target ranking tasks. During the pre-training phase, they use contrastive learning and InfoNCE loss \cite{oord2018representation} to generate user and item embeddings. These embeddings are then used as features during the fine-tuning phase, where a traditional DLR model is trained to enhance the ranking performance. QARM \cite{luo2024qarm} further quantizes the embedding into semantic IDs for downstream training, making these information learnable.  Notably, LUM employs a three-stage pipeline: the first stage focuses solely on next item prediction; the second stage learns to generate user and item embeddings through contrastive learning; the third stage trains DLMs for recall and ranking by using user and item embeddings as input features.

\paragraph{Model-based finetuning}
This category largely falls under the paradigm of end-to-end recommendation, where the pre-training phase learns the ability to predict next items and the finetuning phase then enhances the model’s ranking capability in specific application scenarios using reinforcement learning. Both OneRec \cite{deng2025onerec} and OneSug \cite{guo2025onesug} follow this framework, applied to video recommendation and query suggestion, respectively. EGA-V2 \cite{zheng2025beyond} and EGA-V1 \cite{qiu2025one} are specifically designed for advertising scenarios, achieving notable improvements in end-to-end ad recommendation and ad ranking. 

\subsection{Inference efficiency}
The increased inference latency accompanied by the complex architectures of LLMs presents another challenge that hinders the deployment of GR models in real-world industrial scenarios. Currently, substantial efforts have been dedicated to optimizing decoding speed, primarily focusing on sequence compression, model architecture optimization, and specialized modeling and decoding tricks tailored for the recall and ranking stages.

Compressing the sequence length serves as an effective strategy to fundamentally reduce computational costs. GenRank \cite{huang2025towards} introduces an action-oriented sequence organization framework, which treats items as positional context, halving the input sequence length. DFGR \cite{guo2025action} reduces input sequence length by merging user-item interactions into single tokens through a real-flow and fake-flow. KuaiFormer \cite{liu2024kuaiformer} employs an adaptive item compression mechanism to reduce input sequence length by grouping earlier user interactions into coarsely aggregated representations while retaining fine-grained modeling of recent items, thereby decreasing sequence length without sacrificing recall performance. 

Some work improves decoding efficiency by making slight adjustments to the model architecture. HSTU \cite{zhai2024actions} replaces the standard softmax in attention computation with a pointwise aggregated attention mechanism, reducing the quadratic complexity of self-attention to linear complexity. EGA-V1 \cite{qiu2025one} proposes RecFormer, which introduces a cluster-attention mechanism in the Global Cluster-Former module to replace standard self-attention, reducing computational complexity by dynamically grouping keys/values into semantically coherent clusters via a learnable cluster matrix.

Some tailored tricks are proposed to accelerate the inference speed during the recall and rank stages. To reduce the vocabulary size during recall, some approaches \cite{rajput2023recommender,yang2024unifying,yin2024unleash} attempt to replace item IDs with the hierarchical semantic IDs. These semantic IDs are generated via RQ-VAE, which encodes item content into semantically meaningful token sequences, enabling efficient knowledge sharing across similar items by leveraging hierarchical semantic structure for autoregressive item prediction. RPG \cite{hou2025generating} generates long, unordered semantic IDs in parallel using optimized product quantization, trains with MTP loss to integrate sub-item semantics. To further enhance the inference efficiency of semantic ID-based methods, EGA-V2 \cite{zheng2025beyond} employs Multi-Token Prediction (MTP) inference to enhance scalability and alignment with business objectives. Additionally, URM \cite{jiang2025large} introduces matrix decomposition and probabilistic sampling instead of TopK selection to reduce computational complexity. The key to enhancing inference efficiency in the ranking stage lies in how to efficiently score all candidate items. \citet{zhai2024actions} introduce M-FALCON that processes multiple candidate items in a single forward pass by modifying causal attention masks to ensure mutual invisibility among candidates. Similar to M-FALCON, MTGR \cite{han2025mtgr} introduces a customized masking strategy to prevent information leakage while enabling efficient candidate scoring. 

\subsection{Cold start and world knowledge}
The cold start problem refers to the challenge of generating accurate recommendation outcomes when there is insufficient data, particularly for newly registered users and newly uploaded items~\cite{lam2008addressing,wei2020fast}. 
LLMs offer two primary strategies for mitigating the cold start problem in recommendation systems. (1) Information Augmentation. The goal is to enhance the input data used in recommendations by incorporating new embeddings and knowledge generated by LLMs. For example, SAID~\cite{hu2024enhancing}, proposed by Ant Group, generates item embeddings based on textual information and integrates them into downstream recommendation tasks. Analogously, CSRec~\cite{yang2024common} fuses metadata-based and common sense-based knowledge derived from LLMs as side information to enhance recommendations. (2) Model Reasoning. The central idea is that LLMs can directly produce recommendation results by leveraging patterns learned from large-scale training data. A representative approach is LLM-Rec~\cite{lyu-etal-2024-llm}, which employs carefully designed prompt strategies to derive effective recommendation solutions.

For the above information augmentation and model reasoning approaches, the underlying rationale for the positive impact of LLMs can be attributed to \textit{world knowledge}. Here, world knowledge refers to the extensive contextual and conceptual knowledge inherent in LLMs, which stems from their training on large-scale datasets spanning diverse domains~\cite{zhang2023large}. For instance, Llama 3~\cite{grattafiori2024llama} and Qwen3~\cite{yang2025qwen3} were pre-trained using 15 and 36 trillion multilingual tokens, respectively, with domain diversity. The world knowledge embedded in pre-trained LLMs allows recommendation systems to effectively learn user-item interaction patterns during the cold start stage. Specifically, LC-Rec~\cite{zheng2024adapting} integrates language semantics from Llama with collaborative signals to attain world knowledge and task-specific characteristics in recommendation systems. Notably, recent research has revealed that item representations linearly mapped from language representations in LLMs enhance recommendation performance, demonstrating the value of world knowledge in recommendation systems~\cite{sheng2024language}. Furthermore, the world knowledge within LLMs can be refined by adding external domain-specific knowledge using the Retrieval-Augmented Generation technique~\cite{arslan2024survey}.

The world knowledge in LLMs for recommendation can be obtained from various data sources on users and items: images~\cite{radford2021learning}, videos~\cite{covington2016deep}, and speech~\cite{cui2024recent}. The integration of diverse data sources can be referred to as \textit{multi-modal learning}. To better leverage the multi-modal representations, a potential solution is contrastive learning, such as image-text contrastive loss~\cite{li2021align}, which is used to align image and text representations before representation fusion.

\begin{table}[h]
\caption{\label{multimodal} 
Multi-modal LLMs for recommendation.}
\resizebox{0.48\textwidth}{9.8mm}{
\begin{tabular}{ccc}
\hline
 & Data & Target \\ \hline
InteraRec~\cite{karra2024interarec} & Image, text  & Product  \\
I-LLMRec~\cite{kim2025image}  & Image, text  & Product  \\
NoteLLM-2~\cite{zhang2024notellm} & Image, text  & Note   \\
TALKPLAY~\cite{doh2025talkplay}  & Audio, text  & Music  \\ \hline
\end{tabular}}
\end{table}

The increasing feasibility of multi-modal learning in recommendation systems has been driven by the prosperity of multi-modal LLMs such as CLIP~\cite{radford2021learning}, vision transformer~\cite{kim2021vilt}, and Qwen2-vl~\cite{wang2024qwen2}. Typical multi-modal LLMs for recommendation are summarized in Table~\ref{multimodal}. Specifically, NoteLLM-2~\cite{zhang2024notellm}, developed by Xiaohongshu, utilizes visual information within LLMs to recommend notes to users, resulting in a 6.4\% increase in note exposures. Alternatively, TALKPLAY~\cite{doh2025talkplay} is a multi-modal music recommendation system that encodes audio features, including lyrics and semantic tags, into LLMs to provide music recommendations. TALKPLAY demonstrates superior performance on the Million Playlist Dataset~\cite{chen2018recsys}, which contains cold-start items in its test sets. Furthermore, InteraRec~\cite{karra2024interarec} is an online e-commerce product recommendation method that extracts valuable information from high-frequency web page screenshots. Together, given the standard representation learning frameworks in LLMs, these external multi-modal signals can be easily incorporated into existing LLM-based GRs to handle the cold start problem.

\section{Future Directions}
In this section, we explore promising directions for LLM-based GRs across the following aspects.
\subsection{Model scaling}
Since its observation and proposal, the scaling law has become the theoretical foundation for parameter scaling in large language models.  When it comes to scaling, the traditional DLR has two significant drawbacks: 1) with the scaling of the length of the user behavior sequence, the DLR cannot efficiently process entire user behaviors, which limits the model's performance; 2) scaling incurs approximately linear costs in training and inference with the number of candidates, making the expenses unbearably high \cite{chen2021end,pi2020search,han2025mtgr}. For GRs, recent studies have observed some scaling effects \cite{han2025mtgr,wang2025scaling,huang2025towards}. However, in these works, the model sizes are still limited to a relatively small level, such as 0.x B or 1.x B, and the performance improvements of models at much larger sizes have not been well validated. With a larger model size and longer user behavior sequence, it is an auspicious and challenging direction to train a more powerful generative recommendation model. Considering the requirements of line applications, it is also crucial to explore ways of efficient inference.

\subsection{Data cleaning}
As we all know, the quality of training data has a significant impact on the final performance of large language models. Few works in GRs have investigated how to perform data cleaning in the recommendation domain. Unlike traditional linguistic text corpora used in textual LLMs, training data in recommendation systems comprises not only item IDs but also multi-source side information with multi-modal characteristics. How to handle this heterogeneous side info is still an open question. The training corpora in GRs consist of user behavior sequences, posing unique challenges for quality assessment, as there exists no equivalent of grammaticality evaluation in natural language processing to discern the validity of behavioral sequences. Developing frameworks to evaluate behavioral sequence validity, implement quality-aware data curation through discriminative filtering, and establish dynamic training protocols conditioned on corpus quality represents a principled methodology for substantially improving recommendation performance \cite{huang2024exit}.

\subsection{One model for all}
The core aspiration of LLMs is to achieve a universal architecture capable of accomplishing all diverse language tasks through prompt switching with a single model; recently, remarkable advancements in multimodal large models have further ignited researchers' enthusiasm for developing unified frameworks that support multiple modalities \cite{girdhar2023imagebind,huang2023language,yu2023spae,yang2023teal,zheng2024adapting}. \citet{zhai2024actions} and \citet{deng2025onerec} unify the recall and rank in one GR model. Recently, \citet{jiang2025large} stepped further and proposed that GRs (URM in their work) can function as universal recommendation learners, capable of handling multiple tasks within a unified input-output framework, eliminating the need for specialized model designs. URM can handle multi-scenario recommendation (search included), multi-objective recommendation, long-tail item recommendation, etc. We posit that unifying the input and output, recommendation and search, through generative large models, which deliver customized recommendations by dynamically interpreting user instructions, will emerge as a promising research frontier in next-generation information retrieval.

\section{Conclusions}
In this paper, we have presented a comprehensive survey of LLM-based GRs with a focus on recent advancements. Initially, we outline the general preliminaries and application cases of LLM-based GRs. Subsequently, we introduce the main considerations when LLM-based GRs are applied in real industrial recommendation systems.  Our survey also sheds light on their capabilities across diverse scenarios and promising future directions in this rapidly evolving field. We hope this survey can provide insights for researchers and contribute to the ongoing advancements in the GR domain. 

\section*{Limitations}
In this paper, we embark on a comprehensive exploration of the current LLM-based GRs landscape, presenting a synthesis from diverse perspectives enriched
by our insights. Acknowledging the dynamic nature of GRs, it is plausible that certain aspects may have eluded our scrutiny, and recent advances might not be entirely encapsulated. Given the constraints of the page limits, we are unable to delve into all technical details and have provided concise overviews of the core contributions of mainstream GRs.



\bibliography{custom}




\end{document}